\documentclass [sigconf, noacm]{acmart}

\AtBeginDocument{%
  \providecommand\BibTeX{{%
    \normalfont B\kern-0.5em{\scshape i\kern-0.25em b}\kern-0.8em\TeX}}}

 \acmConference[HRI '24 Companion]{Companion of the ACM/IEEE International
 Conference on Human Robot Interaction (HRI) (HRI ’24 Companion)}{March 11--15, 2024}{Boulder, Colorado}

\setcopyright{none}
\acmDOI{}
\acmISBN{}

\usepackage{hyperref}
\usepackage{graphicx}

\begin{document}

\title[Toward LLM-Powered Social Robots for Supporting Sensitive Disclosures]{Toward LLM-Powered Social Robots for Supporting Sensitive Disclosures of Stigmatized Health Conditions}

\author{Alemitu Bezabih}
\orcid{0000-0001-9603-8537}
\affiliation{%
  \institution{Colorado School of Mines, Department of Computer Science}
  \city{Golden}
  \state{CO}
  \postcode{80401}
  \country{USA}}
\email{alemitubezabih@mines.edu}

\author{Shadi Nourriz}
\orcid{0009-0004-1873-6750}
\affiliation{%
  \institution{Colorado School of Mines, Department of Computer Science}
  \city{Golden}
  \state{CO}
  \country{USA}}
\email{shadinourriz@mines.edu}

\author{C. Estelle Smith}
\orcid{0000-0002-4981-7105}
\affiliation{%
  \institution{Colorado School of Mines, Department of Computer Science}
  \city{Golden}
  \state{Colorado}
  \country{USA}
  \postcode{80401}
}
\email{estellesmith@mines.edu}

\renewcommand{\shortauthors}{Alemitu Bezabih, Shadi Nourriz, and C. Estelle Smith}

\begin{abstract}
  Disclosing sensitive health conditions offers significant benefits at both individual and societal levels. However, patients often face challenges due to concerns about stigma. The use of social robots and chatbots to support sensitive disclosures is gaining traction, especially with the emergence of LLM models. Yet, numerous technical, ethical, privacy, safety, efficacy, and reporting concerns must be carefully addressed in this context. In this position paper, we focus on the example of HIV status disclosure, examining key opportunities, technical considerations, and risks associated with LLM-backed social robotics.
\end{abstract}



\maketitle

\begin{figure*}
  \centering
  \medskip
  \includegraphics [width=\textwidth]{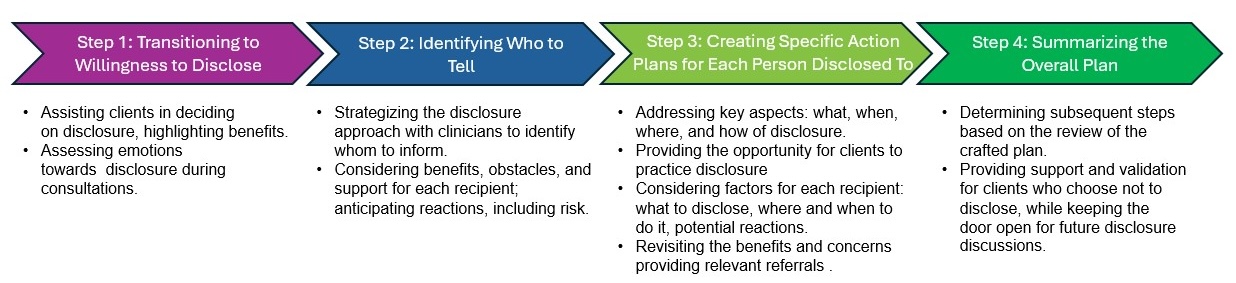}  
  \caption{A four-step self-disclosure model used in clinical practice, for example by the California Prevention Training Center in the UCSF Bixby Center for Global Reproductive Health~\cite{california_ptc_self-disclosure_2024}}
  \label{fig:disclosure-model}
\end{figure*}

\section{Introduction}
For patients struggling with stigmatized physical and mental health conditions, disclosing sensitive information about their health--whether in online contexts such as social media \cite{warner_privacy_2018, andalibi_testing_2018} or in-person social interactions \cite{bezabih_challenges_2023} --poses numerous challenges, including: lack of \textit{a priori} knowledge and skill for how, when, and whom to disclose to; navigating complex interpersonal dynamics prior to, during, and after disclosure; fear of discrimination or social ostracization; and practical concerns like access to healthcare, employment security, and financial stability. However, successful sensitive disclosure can also benefit both individuals and society by fostering empowerment, reducing anxiety, and improving health and quality of life outcomes. Transparency also cultivates trust, facilitates informed decision-making for relationships and healthcare, and fosters a more inclusive community by reducing stigma and promoting empathy. 

The use of social robots in HRI (and chatbots in HCI) to support sensitive disclosures is rapidly gaining research traction \cite{li_tell_2023, yu_i_2019, noguchi_personality_2020, laban_informal_2022, kawasaki_assessing_2020, soleymani_multimodal_2019} ---particularly considering the increasing public availability of LLM models which can readily be implemented as backend models for disclosure systems rather than, for example, traditional Wizard-of-Oz techniques to mimic human-like responses. Yet there are numerous technical, ethical, privacy, safety, efficacy, and reporting concerns that must be carefully discussed in order to ultimately realize the potential of social robots for supporting sensitive disclosures. In this position paper, we consider a case study of disclosure of human immunodeficiency virus (HIV)-positive status, a sexually transmitted infection (STI). Using a 4-step model of how clinicians now navigate HIV disclosure plans with patients, we explore the nuances of this stigmatized physical illness and outline some of its major opportunities, considerations, and risks for social robotics. We posit that future work must make similar considerations for other stigmatized physical and mental illnesses, while also adapting to their unique idiosyncrasies.

\section{Prior Work in Sensitive Disclosures}
Social robots in HRI and conversational agents in HCI are gaining traction toward facilitating sensitive disclosures of challenging personal experiences and emotions. For instance, robotic and chatbot conversations are effective in creating emotionally safe spaces for self-disclosure~\cite{li_tell_2023}, improving and encouraging self-disclosure~\cite{yu_i_2019, noguchi_personality_2020, laban_informal_2022}, and diagnosing and improving mental health~\cite{kawasaki_assessing_2020, soleymani_multimodal_2019}. These HRI/HCI efforts have greatly contributed to designing ways to encourage people to share personal and sensitive information that they might not feel comfortable disclosing to humans. However, several ethical issues remain unresolved---\textit{e.g.,} data \& privacy, informed consent \& autonomy, and loss of empathy~\cite{farhud_ethical_2021}. Moreover, the recent public release of LLMs (\textit{e.g.,} ChatGPT) has greatly attracted HRI/HCI scholarship to integrate LLMs into social robots, exacerbating preexisting risks with new concerns. Therefore, scholars must proactively identify risks in domain-specific designs and contribute to evolving guidelines. Moving toward developing such guidelines, we consider the case of HIV status disclosure via social robots, drawing from an existing 4-step clinical model.

\section{Background: The 4-Step Clinical Model of Disclosure} \label{sec:4-step}
In contexts like HIV/STIs, patients must disclose their HIV status to sexual partners for the sake of health of the latter. Other vital disclosure contexts include parents disclosing their own HIV status to children, or parents disclosing HIV-status to children born with HIV ~\cite{bezabih_challenges_2023, mequanint_bezabih_challenge_2022}. However, many patients struggle with knowing how, when, and with whom to disclose HIV status. Therefore, clinicians engage in ongoing discussions with patients until they develop a disclosure plan for potentially affected individuals. Figure~\ref{fig:disclosure-model} shows a 4-step disclosure model that is now used in clinical practice to guide this process~\cite{california_ptc_self-disclosure_2024}. The steps include: (1) transitioning to willingness to disclose; (2) identifying who to tell; (3) creating specific action plans for each person disclosed to, and (4) summarizing the overall plan. However, going through these steps is challenging given the time constraints and workload of healthcare workers. Importantly, there is also usually limited or no time available during a clinical consult for \textit{practicing} disclosure.

\section{Key Opportunities for Integration of Social Robots in HIV Clinics}
\begin{figure*}[t]
  \centering
  \medskip
  \includegraphics [width=\textwidth]{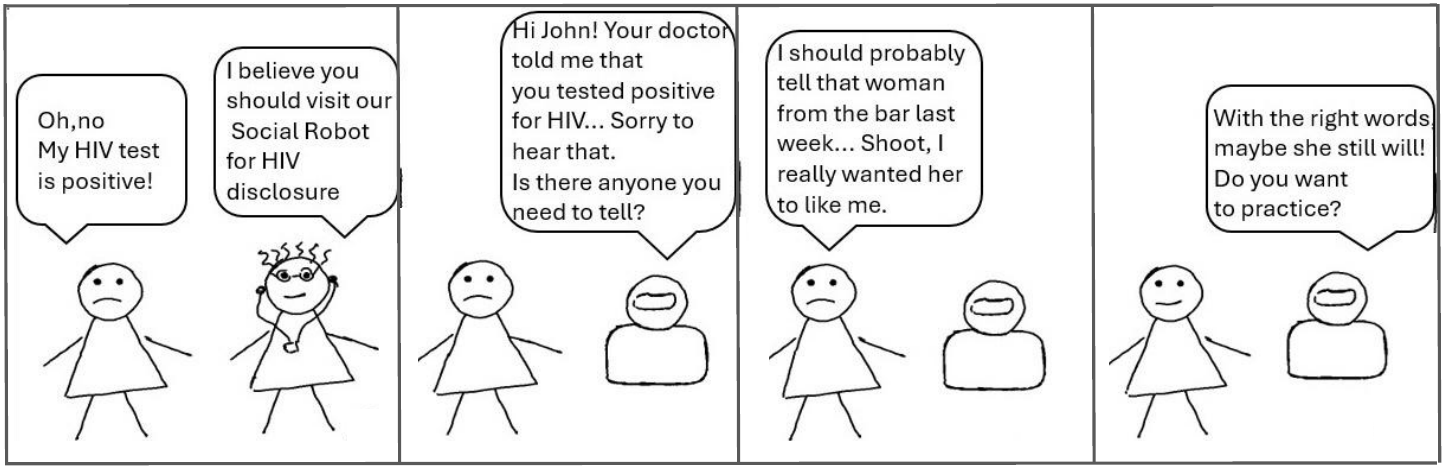}
  \caption{A storyboard depicting a scenario in which a user discovers that he is HIV-positive during a clinical visit. The clinician recommends that he visits with a social robot to practice disclosing his HIV-positive status to important social connections.}
  \label{fig:storyboard}
    \Description{A four-panel storyboard depicting an HIV-positive user, a clinician, and a social robot used for practicing sensitve disclosures.}
\end{figure*}

Social robots powered by LLMs could be helpful to support HIV disclosure and enhance clinical care. For example, in cooperation with clinicians, they could prepare clients for disclosure via some or all of the four steps listed above. Moreover, given the significant lack of clinician time available, a social robot could provide a new 5th step by offering training and practice opportunities: Patients could verbally and repeatedly practice disclosure with the robot, as a stand-in for the intended future human recipient. Informed by prior work in HRI such as~\cite{zlotowski_anthropomorphism_2015}, we hypothesize that such practice may be highly relatable and transferable to real-life conversations; thus brainstorming and practicing sensitive disclosures with a social robot could improve later disclosure to a human. In order to become clinically feasible given clinicians' time constraints and staffing availability, this practice time would likely not be supervised by staff. Figure~\ref{fig:storyboard} depicts a storyboard sketch illustrating the potential for a human patient to practice disclosure with a robot, as described in the hypothetical scenario in Section~\ref{sec:scenario}.

\subsection{An Example Scenario}\label{sec:scenario}
\textit{John, a young man in his twenties, recently found himself in a difficult situation. Having struggled with drug addiction, he made the brave decision to undergo an HIV test. The first test brought temporary relief as it turned out negative. However, a subsequent test revealed the harsh reality - John was HIV positive. Feeling overwhelmed and uncertain, John sought support from his regular clinician. After having a couple of counseling sessions, the clinician helped him to consider disclosing to his circles. To aid this process, the clinician recommended him to a clinic's tech room equipped with a social robot designed to assist individuals in sensitive situations like his.}

\textit{Upon entering the tech room, John was greeted by the social robot, which introduced itself as his supportive companion throughout his disclosure journey. The robot knows everything about John's health profile from his electronic medical record (EMR). The robot began by guiding John through the process of identifying people to disclose his HIV status to, suggesting pre-existing categories such as family, friends, sex partners, co-workers, and providers. Next, the robot helped John brainstorm the benefits and concerns associated with disclosing to each person, allowing him to weigh the potential outcomes carefully. With the robot's assistance, John created a detailed action plan for each individual on his list, outlining how and when he would disclose his status to them. Through a UI/UX interface, John and the robot crafted a disclosure plan and attached it to the EMR.} 

\textit{To prepare John for the challenging conversations ahead, the robot provided a safe space for him to practice disclosure, allowing him to say his thoughts out loud and receive feedback and suggestions for improvement. The robot's non-judgmental presence and supportive feedback helped John build confidence in expressing himself. As the session drew to a close, the robot summarized the overall plan, revisiting specific action plans and key issues discussed during the session. Additionally, the robot reminded John of the importance of confidentiality and the potential risks of breaches. Feeling empowered and supported by the guidance of the social robot, John left the tech room with a sense of clarity and determination to navigate the journey of HIV disclosure with confidence and resilience.}

\section{Technical \& Design Considerations}

\paragraph{\textbf{Which steps are [in]appropriate for a social robot to manage?}} Of the steps outlined, it is not immediately clear which steps are best supported by a social robot. Which aspects of care are best delivered by humans, and which by robots? For instance, are clinicians or robots more persuasive in convincing a human to disclose in the first place? Are clinicians or robots more effective in suggesting targets of disclosure, plans to disclose, and summaries of the plan? Research should first ascertain the degree to which it is both clinically feasible and desired by patients for clinicians or robots to administer each of the needed steps.

\paragraph{\textbf{How might LLMs be technically integrated?}}
In a clinical context, information can be collected and shared via screening tools (\textit{e.g.,} intake surveys or interviews), online messaging with clinicians through electronic portals, in-person discussions with clinicians during appointments, and/or electronic health records (EHRs). In order to effectively interact with patients, a social robot will either need to ingest previously collected data from such source(s), or collect information itself during interaction with a patient. Moreover, its conversations must be structured appropriately to meet clinical aims. Therefore a robot must be architectured with consideration for how and when use of an LLM is appropriate in the workflow. Some steps may be best served by structured, consistent queries (\textit{e.g.,} intake questions, determining who to disclose to, or summarizing a plan) without need for an LLM. In other cases, natural language may be essential (\textit{e.g.,} persuading to disclose, or practicing disclosures) for a fluid and effective user experience. The architecture of the robot must therefore be capable of reasoning about which step of the process it is involved in, use the correct LLM or non-LLM based strategy for that step, have appropriately trained and scoped LLM-models available for LLM-based steps, and have access to the appropriate PII for each patient. Thus, questions arise related to good training materials for each LLM-based step, as well as technical access to electronic discussion portals, EHRs, or in-person dialogues.

\paragraph{\textbf{Reporting concerns}}
In steps that eventually do involve LLMs, it is essential that researchers proceed cautiously, and record, review and report on LLM performance. Clinicians should be involved in the review of interactions, and comparisons made between the language clinicians would choose to use at each step, and the language generated by an LLM. Later technical development should be taken to adjust LLM-generated utterances toward clinician recommendations. Patient feedback is also essential. For instance, preliminary data suggest that LLMs can be perceived as more empathetic than clinicians \cite{ayers_comparing_2023}; conversely, if LLMs can take care of some steps, this may enable more time for clinicians to express more empathy \cite{topol_machines_2023}. Researchers should report on: how LLMs were trained at each step; the performance of LLMs according to clinicians and patients; and what steps can be taken to improve performance.

\section{Risks \& Ethics in HIV Disclosure}\

\paragraph{\textbf{Safety and efficacy}}
Issues may arise concerning the quality of LLM-generated robot conversations, such as misinformation. An LLM-backed social robot could potentially offer non-factual information, including deliberate disinformation (\textit{e.g.,} stemming from malicious attacks on LLMs), unintentional inaccuracies or biases inherent in the LLM's training data, or subtle misstatements that reinforce stigma. For instance, there is a risk that a robot may suggest an ineffective strategy or an inappropriate target for disclosure. Consequently, questions emerge regarding how to filter information originating from the LLM to mitigate possible harms. Furthermore, given the highly sensitive nature of HIV, individuals living with the condition often harbor traumatic and stigmatizing experiences. Engaging in a conversation with a robot may inadvertently trigger these negative experiences, leading to emotional distress. While such triggers may also occur in human interactions, clinicians are equipped to provide emotional support in such instances. Thus, the challenge lies in detecting and effectively managing these triggers when they arise during interactions with a social robot. In conclusion, there is a pressing need for mechanisms to evaluate the effectiveness of robotic interventions in comparison to standard care protocols.  

\paragraph{\textbf{Privacy \& HIPAA}}
The concern of privacy, particularly in sensitive contexts, raises critical questions regarding the integration of social robots with any clinical data, EHRs, or Personal Identifiable Information (PII)~\cite{grande_health_2020}. With the stringent regulations outlined in HIPAA~\cite{hhs_hipaa_2024}, concerns arise about whether a robot \textit{should} have direct access to such data. One potential approach involves implementing additional privacy-preserving layers to ensure the protection of PII--\textit{e.g.,} stripping PII from prompts before they are processed by LLMs, and then reintroducing it at the conversational layer as needed. By incorporating such measures, we can strike a balance between leveraging the capabilities of social robots in healthcare contexts while safeguarding patient privacy following HIPAA regulations. Implementing additional privacy-preserving layers may increase computational overhead and processing time, potentially slowing down system response. Optimization strategies, such as efficient algorithms and hardware acceleration, may be necessary to mitigate these efficiency and speed implications.

\paragraph{\textbf{Ethical and social concerns}} 
The introduction of robots into sensitive domains raises significant ethical and social concerns regarding their appropriateness. For instance, clinicians adhere to strict codes of ethics (\textit{e.g.,} The Code of Ethics for Nurses~\cite{american_nurses_association_code_2017}), yet how to hold robots accountable remains an open question. Moreover, ethics of care emphasize the importance of relationships, empathy, and compassion in human-centered care of patients~\cite{gilligan_different_1993}, however robots and LLMs are not sentient beings capable of genuine care---despite their appearance or mimicry of sentience. For instance, the following is a quote from a chaplain participant in \cite{smith_what_2021}: \textit{``God help us if it’s just robots taking care of people at end of life.''} In the extreme case of terminal illness, this quote evokes a philosophical dread for the loss of soulful, loving connection that is vital to human wellbeing, meaning-making, and purpose. Limits should be limits imposed on the degree to which robots can replace humans in clinical care, and these must be considered with excessive caution and forethought. Establishing frameworks for robot accountability through regulations, standards of care, and oversight mechanisms, is essential to ensure their responsible use and to mitigate potential risks to individuals and society at large. 

\section{Proposed Experimental Design}
In our future work, we aim to investigate these questions situated within an HIV clinic at our local Children's Hospital. In our first phase of work, we plan to conduct design work in collaboration with HIV clinicians. We will: (1) investigate which steps clinicians recommend be integrated with a robot and explore the feasibility and acceptability ramifications of these steps; (2) design and prototype the robot's cognitive architecture; (3) determine the technical requirements required to protect patient privacy and safety. 

Building upon prior work~\cite{bezabih_challenges_2023}, our prototype design will focus on parents disclosing their HIV status to children. Here, we need to highlight that while the four-step model described in section~\ref{sec:4-step} assumes disclosure as a one-off event, this approach may not be appropriate for children. Children have a limited understanding of illness processes which advances with age~\cite{bibace_development_1980}. This implies that HIV information needs to be structured around stages of cognitive development and should be delivered progressively following children's capacities to understand illness~\cite{lesch_paediatric_2007}. In consultation with HIV clinicians, we will determine a specific population of children at a particular developmental stage to work with for this study. For the technical development of the prototype, we will utilize a child-faced Furhat robot integrated with a fine-tuned GPT-4 LLM model. To mitigate the safety and efficacy concerns presented above, we plan to train our model on a customized dataset consisting of HIV and disclosure resources including peer-reviewed HIV research, disclosure guidelines, and governmental policies. We will experiment with a variety of prompt engineering strategies to balance an empathetic, caring, and understanding tone against a prescriptive or persuasive clinical tone. 

Following this careful stage of prototyping and co-design with clinicians, we propose to conduct an experiment in which we compare patients' perceptions of and benefits derived from disembodied LLM-based conversational agents (i.e. a chatbot using our LLM model) \textit{v.s.} similar agents embodied within the physical Furhat robot. We plan to conduct a one-month trial of our prototypes with 30 parents living with HIV (15 to trial the disembodied UI/UX version, and 15 to trial the robot format), as a supplemental service to standard disclosure support. To demonstrate whether there is a meaningful advantage to using a robot over a chat interface, a pre-post disclosure readiness assessment will be conducted based on the six stages of decision readiness to disclose their HIV-positive status to children as described in~\cite{laschober_decision_2019}. 

\section{Conclusion}
The disclosure of sensitive health conditions holds immense benefits but is often hindered by stigma-related challenges. The rising use of social robots and chatbots, particularly with advancements in LLM models, may help to support sensitive disclosures. However, the integration of these technologies raises numerous technical, ethical, privacy, safety, efficacy, and reporting concerns. Focusing on HIV status disclosure as an example, our position paper explores opportunities, technical considerations, and risks associated with LLM-backed social robotics in healthcare. We explore which aspects of the disclosure process are suitable for social robots, technical integration of LLMs, safety, efficacy, privacy, HIPAA compliance, and broader ethical and social implications. Through this exploration, we aim to offer insights into responsible and ethical deployment of social robots in sensitive healthcare contexts.

\bibliographystyle {acm}
\bibliography{disclosure-robotics}
\end{document}